\documentclass{article}
\usepackage{graphicx}
\usepackage{amsmath}
\usepackage{amsfonts}
\usepackage{amssymb}

\begin{document}

\title{A Statistical Analysis of Log-Periodic Precursors to Financial Crashes}
\author{James A.\ Feigenbaum\\Tippie College of Business\\Department of Economics\\University of Iowa\\Iowa City, 55242}
\maketitle
\begin{abstract}
Motivated by the hypothesis that financial crashes are macroscopic examples of
critical phenomena associated with a discrete scaling symmetry, we reconsider
the evidence of log-periodic precursors to financial crash\-es and test the
prediction that log-periodic oscillations in a financial index are embedded in
the mean function of this index. \ In particular, we examine the first
differences of the logarithm of the S\&P 500 prior to the October 87 crash and
find the log-periodic component of this time series is not statistically
significant if we exclude the last year of data before the crash. \ We also
examine the claim that two separate mechanisms are needed to explain the
frequency distribution of draw downs in the S\&P 500 and find the evidence
supporting this claim to be unconvincing.
\end{abstract}

\newpage

\section{Introduction\bigskip}

\begin{figure}
[ptb]
\begin{center}
\includegraphics[
natheight = 4.264in,
natwidth = 8.556in,
height=150.8125pt,
width=299.1875pt
]%
{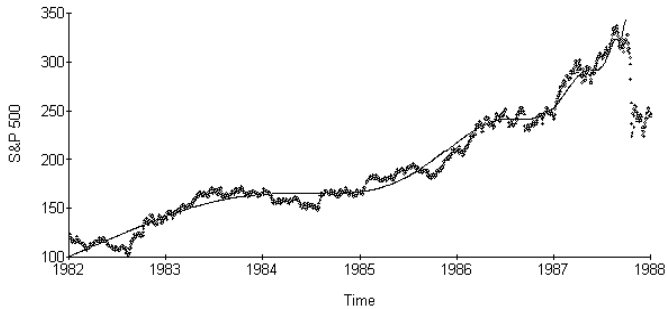}%
\caption{The S\&P\ 500 from 1982 to 1988 and a fit according to Eq. (\ref{lp})
with $A=380.4$, $B=-10.8$, $C=-.0743$, $t_{c}=10/20/87$, $\alpha=.422$,
$\omega=5.965$, and $\phi=.124$.}%
\label{abs index}%
\end{center}
\end{figure}

\qquad Several authors in the physics community (including \cite{BJS 96},
\cite{FF 96}, \cite{FF 98}, \cite{GY 98}, \cite{JS 99a}, \cite{SJB 96},
\cite{Vandewalle 98a}, and \cite{Vandewalle 98b}) have written about the
appearance of log-periodic oscillations in stock market indices during the
months and years leading up to financial crashes. \ By a log-periodic
oscillation, we mean that a variable exhibits a functional form such as
\begin{equation}
y(x)=A+Bx^{\alpha}\left[  1+C\cos(\omega\ln x+\phi)\right]  \text{.}\label{lp}%
\end{equation}
The cosine term in this expression produces oscillations with a period that
grows or shrinks exponentially. \ As an example, if we let $x=t_{c}-t$, where
$t_{c}$ is the time of the well-known crash of October 19, 1987, then we can
see this pattern in the Standard and Poor's 500 (S\&P~500) during the five
years prior to the crash as shown in Fig.~\ref{abs index}.

\qquad Such log-periodic oscillations are of interest to physicists because
log-periodicity is the signature of a spatial environment with a discrete
scaling symmetry. \ The observation of log-periodic precursors has been
interpreted as evidence that the subsequent financial crash can be viewed as a
critical point analogous to a phase transition in a more traditional physics
environment. \ Heuristically, this picture of a transition to an ordered state
is certainly consistent with what happens during a stock market crash, when
virtually all investors will be in agreement that the price is dropping and
that they should sell, in contrast to the normal disordered state where the
price remains fairly stable because there is a buyer for every seller. \ The
theoretical issue is how far this analogy can properly be taken.%

\qquad One can envision a discrete scaling symmetry arising in financial
markets if one models investors as forming networks of friends and contacts.
\ If every agent has $n$ contacts, it is indeed possible to construct a
network which exhibits a scaling symmetry. \ Johansen et al. \cite{JLS 98}
propose a model of imitative investors, living in such a network, which is
essentially an Ising spin model. \ A spin-up constituent represents a buyer, a
spin-down constituent represents a seller, and neighboring spins interact in a
manner that encourages them to align in the same direction. \ Johansen et al.
respond to the criticism that a rational agent could exploit log-periodic
fluctuations to his financial advantage, and thereby acquire all the wealth in
the economy, by requiring that stock prices obey a martingale condition that
would rule out arbitrage opportunities. \ How the collective behavior of the
imitative investors translates into a price that satisfies this martingale
condition is left unmodeled. \ Regardless of the underlying details, this
martingale condition is a very strong restriction on the behavior of prices,
and testing this condition is the primary focus of the present paper.

\qquad A major deficiency of the vast body of previous work on log-periodic
precursors is that most of the analysis has been qualitative rather than
statistical in nature. \ The emphasis has been placed on establishing the
existence of log-periodic fluctuations, usually through curve-fitting or more
recently through the use of Lomb periodograms (\cite{JLS 99}). \ We should
note that we take it as a given in this paper that such methods are valid.
\ We do not dispute the existence of log-periodic fluctuations like those
pictured in Fig. \ref{abs index}. \ What remains controversial is why these
fluctuations exist. \ Are they indeed part of a collective process leading up
to a phase transition?\ \ Or can they be more simply explained as an
accidental feature of the stochastic processes that drive financial indices?
\ Only statistical methods have the power to distinguish these possibilities.

\qquad A small amount of work has already been done in this vein. \ Feigenbaum
and Freund \cite{FF 96} modeled the stock market as a random walk and found
that the average time between a log-periodic spell and the next financial
crash was significantly longer in real-world data than in Monte Carlo
simulations. \ Johansen et al. (\cite{JLS 98}, \cite{JLS 99}) report similar
results from an experiment where they modeled the stock market as a GARCH
process. \ While these are positive results with respect to the hypothesis of
crashes as critical points, they are both strongly tied to a specific model of
the underlying stochastic process, what statisticians term the data generating
process (DGP). \ One cannot extrapolate from these two results to conclude
that no DGP\ can account for log-periodicity. \ The null hypothesis that
log-periodic spells are the casual products of random fluctuations.has
\emph{not} been rejected.

\qquad While most studies of log-periodicity have focused on prices, the
present paper applies statistical hypothesis testing to the question of
whether the first differences of a price process exhibiting log-periodic
fluctuations are also log-periodic. \ This question is of interest because any
model in which log-periodic price fluctuations reflect corresponding
fluctuations in the mean of the price process will require that the mean of
the first differences should also behave log-periodically. \ In particular,
the model of Johansen et al. \cite{JLS 98} requires this. \ Moreover, it is
well-known among econometricians that regression estimators are inconsistent
when applied to a serially-correlated time series like a stock price. \ First
differencing the stock price can substantially eliminate this serial
correlation and may eliminate effects produced by it. \ Consequently, the
absence of log-periodic fluctuations from the first differences of a price
process suggests (although certainly does not prove) that log-periodicity in
the integrated price process may be due to serial correlation and random effects.

\qquad Examining the first differences for the logarithm of the S\&P 500 from
1980 to 1987, we find that the log-periodic component of the Sornette-Johansen
specification is statistically significant if we include this entire period in
the fit. \ However, if we exclude the last year, the log-periodic component is
no longer statistically significant. \ Furthermore, the best fit in this
truncated data set predicts a crash in June of 1986, shortly after the data
set ends but well before the actual crash. \ Presumably the log-periodic fit
is coupling to some structure in the data just prior to the crash which does
not characterize the whole time series. \ Thus we conclude that
log-periodicity is either negligible or not present in these first
differences, a negative result for the model of Johansen et al. \cite{JLS 98}.

\qquad A second claim that we consider was put forward by Johansen and
Sornette \cite{JS 99b} and pertains to the distribution of drawdowns, where a
drawdown is the accumulated drop in an index from a local maximum to the next
local minimum. \ Assuming that the fat-tailed distribution of daily returns
falls off exponentially in the limit of large negative returns, they derive
that the distribution of drawdowns should also be exponential. \ Empirically,
they find that the frequency distribution of drawdowns over the past century
does indeed look exponential except for three outliers, the 29 and 87 crashes
and a crash during World War I. \ They infer from this that a different
mechanism must be responsible for these outliers and suggest that this
mechanism is also responsible for producing log-periodic oscillations. \ In
principle, however, one can examine any set of data with one dependent
variable and say that the graph looks exponential except for some outliers.
\ In order to make this a testable claim, one has to consider whether there
are statistically significant deviations from the exponential curve. \ In
fact, statistically significant deviations do occur, and we conclude that this
evidence of a special crash mechanism is unconvincing.

\qquad The paper proceeds as follows. \ In Section \ref{SJP}, we review the
basic elements of the Sornette-Johansen framework and note some theoretical
criticisms. \ Since the initial papers on this subject appeared, Sornette and
Johansen \cite{SJ 97} have developed a more sophisticated specification for
observed log-periodic oscillations, treating the logarithm of the price index
as the dependent variable rather than the price itself. \ We will see in
Section \ref{logspec} that the logarithmic specification does fit the
S\&P\ 500 prior to the 87 crash. \ It also fits to the S\&P 500 prior to
downturns in 1974, 1997, 1998, and an otherwise unremarkable downturn of less
than 1\% in 1985. \ This last finding raises the spectre that perhaps
log-periodic behavior is not especially noteworthy and is not inherently
related to crashes. \ In Section \ref{FD}, we examine the first differences of
the logarithm of the S\&P\ 500 during the period of 1980 to 1987 and test the
statistical significance of the log-periodic component of
the\ Sornette-Johansen specification. \ In Section \ref{DDD}, we consider the
distribution of drawdowns over the period from 1962 to 1998 and examine
whether the 87 crash deserves classification as an outlier. \ Finally, in
Section \ref{CR}, we offer some concluding remarks.\bigskip

\section{The Sornette-Johansen Paradigm\label{SJP}\bigskip}

\qquad Johansen et al. \cite{JLS 98} model the behavior of a speculative stock
which pays no dividends. \ They allow for the possibility that at least one
rational, risk neutral agent with rational expectations, who does not discount
the future, exists in their economy. \ Given this assumption, the stock price
$p(t)$ must follow a martingale. \ For all $t^{\prime}>t$,
\begin{equation}
E_{t}\left[  p(t^{\prime})\right]  =p(t),\label{martingale}%
\end{equation}
where $E_{t}$ denotes the expectation operator conditional on all information
available at $t$. \ Strictly speaking, the $E_{t}$ operator in Eq.
(\ref{martingale}) would refer to an expectation based on the beliefs of the
hypothesized agent. \ When we say the agent has rational expectations, we mean
the agent's beliefs in fact correspond to reality, so $E_{t}$ is also the
objective expectation operator.

\qquad In a market equilibrium where agents behave optimally according to
their preferences, Eq.~(\ref{martingale}) is an endogenous no-arbitrage
condition that must hold for markets to clear. \ If $E_{t}[p(t^{\prime
})]>p(t)$, a risk-neutral agent will find it profitable to hold an infinite
amount of the stock. \ Consequently, rational agents will bid up the price at
$t$ until Eq. (\ref{martingale}) is satisfied. \ Similarly, if $E_{t}%
[p(t^{\prime})]<p(t)$, rational agents will find it profitable to sell short
an infinite amount of the stock. \ Since the supply of stock is finite,
neither of these two possibilities is consistent with equilibrium.

\qquad Since the stock pays no dividends, its fundamental value (the expected
present value of the stream of future dividends) is zero, so any positive
price constitutes a ``bubble''. \ In a world of perfect certainty, for
$t^{\prime}=t+1$, Eq. (\ref{martingale}) would translate to the trivial
difference equation
\begin{equation}
p_{t+1}=p_{t}\text{,}\label{triv diff}%
\end{equation}
and it would follow by mathematical induction that $p_{t}$ must be a constant.
\ If we assume that bubbles cannot be sustained forever, then a transversality
condition (essentially the boundary condition at the end of time) will hold:
\begin{equation}
\lim_{t\rightarrow\infty}p(t)=0.\label{cer limit}%
\end{equation}
In that case, we will have $p_{t}=0$ for all $t$. \ Given the transverality
condition, bubbles cannot arise under perfect certainty. \ With uncertainty,
however, Eq.~(\ref{martingale}) is no longer a deterministic difference
equation and will have an infinite number of solutions that satisfy the
generalized transverality condition
\begin{equation}
\lim_{t^{\prime}\rightarrow\infty}E_{t}[p(t^{\prime})]=0\text{.}%
\label{uncer lim}%
\end{equation}
(See Chapter 5 of \cite{BF 89} for a review of bubble solutions.)

\qquad Consequently, Johansen et al. consider a price process which satisfies
Eq.~(\ref{martingale}) but also introduces the possibility of a crash. \ Let
$j$ denote a random variable that equals 0 before the crash and 1 afterwards.
\ Denote the continuous distribution function (cdf) of the crash time as
$Q(t)$ and the corresponding probability density function (pdf) as
$q(t)=\dot{Q}(t)$. \ The hazard rate, the probability per unit time that the
crash will happen in the next instant if it has not already happened, will
then be
\begin{equation}
h(t)=\frac{q(t)}{1-Q(t)}.\label{hazard}%
\end{equation}
If we assume that the crash involves a downturn of a fixed percentage
$\kappa\in(0,1)$ of the price, then the price process can be described as
\begin{equation}
\frac{dp}{p(t)}=\mu(t)dt+\varepsilon(t)-\kappa p(t)dj,\label{sde}%
\end{equation}
where $\varepsilon(t)$ is a mean-zero noise term. \ The time-dependent drift
$\mu(t)$ is then determined by Eq. (\ref{martingale}):
\begin{equation}
E_{t}[dp]=\mu(t)p(t)dt-\kappa p(t)h(t)dt=0\text{.}%
\end{equation}
Thus
\begin{equation}
\mu(t)=\kappa h(t).
\end{equation}
Disregarding the noise term (whose contribution may be computed using Ito
calculus), Eq. (\ref{sde}) has the solution
\begin{equation}
\ln\left[  \frac{p(t)}{p(t_{0})}\right]  =\kappa\int_{t_{0}}^{t}h(t^{\prime
})dt^{\prime}\label{price}%
\end{equation}
prior to the crash. \ The connection of this theory to physics lies in the
determination of the hazard rate function $h(t)$.

\qquad In addition to any rational agents in the economy, Johansen et al.
posit the existence of a large number of irrational noise traders, and it is
these noise traders who are responsible for any dynamics leading up to a
crash. \ They construct a ``microscopic'' model in which the noise traders are
imitative investors who reside on a local interaction network. \ Neighbors of
an agent on this network can be viewed as the agent's friends or contacts, and
an agent will incorporate the views of his neighbors regarding the stock into
his own view.

\qquad At time $t$, each agent $i$ is assumed to be in one of two states: \ a
bullish state $s_{it}=+1$ or a bearish state $s_{it}=-1$. \ If we denote the
set of nearest neighbors as $N(i)$, given the states of all his neighbors at
$t$, $i$'s state at $t+1$ will be determined by
\begin{equation}
s_{i,t+1}=\operatorname{sgn}\left(  K\sum_{j\in N(i)}s_{jt}+\varepsilon
_{i,t+1}\right)  ,
\end{equation}
where $K>0$ is a constant and $\varepsilon_{it}$ is an i.i.d. random variable
with $E[\varepsilon_{it}]=0$. \ The sign function $\operatorname{sgn}(x)$
equals 1 for positive $x$ and $-1$ for negative $x$. \ If $i$'s neighbors are
preponderantly bullish, $i$ will likely be bullish also, and conversely if his
neighbors are bearish. \ A crash occurs when virtually all agents are in the
bearish state. \ Johansen et al. are able to show that the hazard rate
function $h(t)$ will be determined by $K$, the distribution of the
$\varepsilon_{it}$, and the structure of the network. \ In particular, if the
network exhibits a discrete scaling symmetry, then the hazard rate function
will have a log-periodic form like Eq. (\ref{lp}), and Eq. (\ref{price}) then
implies that the stock price will also behave log-periodically.

\qquad One criticism of this model made by Ilinski \cite{Ilinski 99} is that
Eq. (\ref{martingale}) will not hold if agents are risk-averse, although the
modification he suggests is not precisely correct. \ If agents allocate their
investments at time $t$ so as to maximize
\begin{equation}
E_{t}\left[  \int_{t}^{\infty}e^{-\delta(\tau-t)}u(c(\tau))d\tau\right]
,\label{U}%
\end{equation}
where $c(t)$ is consumption at time $t$, $u(\cdot)$ is the period utility
function (which measures the degree of satisfaction or happiness that an agent
derives from a given level of consumption), and $\delta$ is the rate at which
agents discount utility from future consumption, then it can be shown that in
equilibrium the price of a stock must obey
\begin{equation}
p(t)=e^{-\delta(t^{\prime}-t)}E_{t}\left[  \frac{u^{\prime}(c(t^{\prime}%
))}{u^{\prime}(c(t))}p(t^{\prime})\right] \label{gen mar}%
\end{equation}
for any $t^{\prime}>t$. \ Note that, in addition to $p(t^{\prime})$,
$c(t^{\prime})$ will also be a random variable, so Eq.~(\ref{gen mar}) cannot
be simplified to an expression of the form
\begin{equation}
p(t)=\nu E_{t}\left[  p(t^{\prime})\right]  ,\label{constant kernel}%
\end{equation}
where $\nu$ is a constant. \ Instead, one typically writes Eq. (\ref{gen
mar})\ as
\begin{equation}
p(t)=E_{t}[\rho(t^{\prime})p(t^{\prime})],\label{gen mar2}%
\end{equation}
where
\begin{equation}
\rho(t^{\prime})=e^{-\delta(t^{\prime}-t)}\frac{u^{\prime}(c(t^{\prime}%
))}{u^{\prime}(c(t))}%
\end{equation}
is often called the pricing kernel. \ Eq. (\ref{gen mar2}) can still be viewed
as describing a martingale process because the pricing kernel can be
interpreted as a Radon-Nikodym derivative which transforms the probability
distribution (\cite{HK 79}). \ However, this transformed probability
distribution will no longer correspond to the objective probability
distribution, even under the rational expectations hypothesis.

\qquad In \cite{JLS 99}, Johansen et al. respond to Ilinksi's criticism, but
their rebuttal applies only to the martingale condition of Eq. (\ref{constant
kernel}) and not to the more general condition of Eq. (\ref{gen mar2}).
\ Their argument is that, since $\nu$ in (\ref{constant kernel}) is a finite
constant, it cannot have any impact on the log-periodicity of $E[p(t^{\prime
})]$, and this would be correct if the pricing kernel was a constant.
\ However, the pricing kernel in general is not a constant and is not even a
function of the price. \ Thus, it is not obvious that variations in the
pricing kernel could not counteract variations in the price in such a way as
to disrupt any log-periodicity.

\qquad Moreover, the determination of the pricing kernel is a highly
non-trivial problem if there exists a multiplicity of rational agents who are
not identical. \ Since Johansen et al. do not model how the collective
behavior of the noise traders is translated into a price, it is not possible
for us to even begin such a computation. \ But corresponding to every rational
agent, there would be an equation like. (\ref{gen mar2}) which would have to
be satisfied, and it is not clear that a log-periodic price trend could be
made compatible with each and every one of these martingale conditions.

\qquad Nevertheless, if we ignore the possibility of risk-averse agents, we do
have a workable model, and we will see in Section \ref{FD} whether Eq.
(\ref{price}) with a log-periodic hazard rate function is consistent with the
behavior of the S\&P 500 prior to the 1987 crash.\bigskip

\section{The Logarithmic Specification\label{logspec}\bigskip}

\qquad Although early investigations of log-periodic fluctuations in stock
market indices prior to financial crashes focused on the specification of Eq.
(\ref{lp}), Sornette and Johansen \cite{SJ 97} introduced an alternative
specification in which it is the logarithm of the price rather than the
absolute price which fluctuates log-periodically. \ This logarithmic
specification is preferable to Eq. (\ref{lp}) because it is widely assumed
that investors are primarily concerned with relative changes in stock prices
rather than absolute changes\footnote{This focus on relative changes can be
understood if investors have constant relative risk aversion (CRRA)
preferences (meaning that $u(c)$ in Eq. (\ref{U}) is either $\ln c$ or
$c^{\gamma}$ for $\gamma\leq1$).}. It is also more straightforward to compare
the logarithmic specification to the once popular hypothesis that the
logarithm of a stock price will follow a random walk with drift.

\qquad Let $t_{0}$ be a time prior to the crash at which the stock price has
begun to fluctuate log-periodically. \ We define
\begin{equation}
f_{1}(t)=\frac{(t_{c}-t)^{\beta}}{\sqrt{1+\left(  \frac{t_{c}-t}{\Delta_{t}%
}\right)  ^{2\beta}}},
\end{equation}
and
\begin{equation}
f_{2}(t)=f_{1}(t)\cos\left[  \omega\ln(t_{c}-t)+\frac{\Delta_{\omega}}{2\beta
}\ln\left(  1+\left(  \frac{t_{c}-t}{\Delta_{t}}\right)  ^{2\beta}\right)
+\phi\right]  .
\end{equation}
Then for $t_{0}<t<t_{c}$,
\begin{equation}
E_{t_{0}}\left[  \ln\left(  p(t)\right)  |j(t)=0\right]  =A+Bf_{1}%
(t)+Cf_{2}(t),\label{logps}%
\end{equation}
where $E_{t_{0}}$ is the objective expectation operator at $t_{0}$ and
$j(t)=0$ if no crash has occurred at or prior to $t$. This specification has
nine free parameters: $\ A$, $B$, $C$, $\omega$, $\beta$, $t_{c}$, $\phi$,
$\Delta_{t}$ $(>0)$, and $\Delta_{\omega}$. \ Note that in the
Sornette-Johansen paradigm, $t_{c}$ does not necessarily correspond to the
time of the crash per se (which we will denote by $t^{\ast}$) but to a
critical time where the probability of a crash becomes very large. \ In terms
of the microscopic model in \cite{JLS 98},
\begin{equation}
\beta=\frac{\eta-2}{\eta-1}\text{,}%
\end{equation}
where $\eta$ is the number of nearest neighbors of each agent. \ Consequently,
if their model is correct, $0<\beta<1$. \ Sornette and Johansen also suggest
making this restriction so that $p(t)$ does not become unbounded as
$t\rightarrow t_{c}$. \ However since log-periodic behavior typically breaks
down as $t$ gets close to $t_{c}$, it has become standard practice to
disregard data points very close to the observed crash, in which case Eq.
(\ref{logps}) will be bounded in the time span of interest even for $\beta<0$.

\qquad The canonical example of a log-periodic precursor can be seen during
the years prior to the October 1987 crash as shown in Fig. \ref{abs index}, so
we first consider the closing value of the S\&P\ 500 for each day between
January 1, 1980 and September 30, 1987\footnote{All data on the S\&P 500 index
were obtained from the Center for Research on Securities Prices (CRSP).}.
\ Suppose there are $n$ trading days in this period. \ Given a choice of
parameters $A$, $B$, $C$, $\beta$, $\omega$, $t_{c}$, $\phi$, $\Delta_{t}$,
and $\Delta_{\omega}$, we define an $n$-dimensional residual vector $e$ by
\begin{equation}
e_{i}=\ln(p(t_{i}))-A+Bf_{1}(t_{i})+Cf_{2}(t_{i}),
\end{equation}
where $t_{i}$ is the time of the $i$th trading day. \ The standard practice
for estimating the parameters of Eq.~(\ref{logps}) is to use the least squares
estimation procedure, choosing those values which minimize the square of the
residual vector $e^{T}e$. \ Econometricians define the coefficient of
determination
\begin{equation}
R^{2}=1-\frac{e^{T}e}{\sum_{i=1}^{n}(y_{i}-\overline{y})^{2}},
\end{equation}
where $y_{i}=\ln(p(t_{i}))$ and
\begin{equation}
\overline{y}=\frac{1}{n}\sum_{i=1}^{n}y_{i}%
\end{equation}
is the sample mean of $\ln(p(t_{i}))$. \ The coefficient of determination is
often interpreted as the ratio of the amount of variation in the dependent
variable explained by a model to the total amount of variation in that
dependent variable. \ The closer $R^{2}$ is to unity, the more variation in
the dependent variable is explained by the model. \ Note however that $R^{2}$
is not invariant to linear transformations of the $y_{i} $, so one should not
put too much emphasis on its value. \ Notwithstanding, one could equally well
characterize the least squares procedure as maximizing the coefficient of determination.

\qquad Eq. (\ref{logps}) is linear in the parameters $A$, $B$, and $C$ and
nonlinear in the remaining parameters. \ We use the downhill simplex method
(see Chapter 10.4 of \cite{NRC}) to minimize $e^{T}e$ as a function of $\beta
$, $\omega$, $\phi$, $t_{c}$, $\Delta_{t}$, and $\Delta_{\omega}$. \ For each
choice of these nonlinear parameters, we obtain the best values of the linear
parameters $A$, $B$, and $C$ using ordinary least squares (OLS) methods.

\qquad The square of the residual vector $e^{T}e$ is not a well-behaved
concave (down) function, so the local minimum obtained by the downhill simplex
method will depend on the initial starting point fed to the routine. \ Indeed,
as was noted in \cite{FF 96}, one typically finds many local minima with
fairly similar values of $e^{T}e$. \ Nonlinear least squares (NLLS) estimation
requires that as the sample size goes to infinity, $n^{-1}e^{T}e$ should
converge to a function with a unique global minimum in some chosen parameter
space (see \cite{Amemiya 85}). \ However, NLLS theory does not require that
$e^{T}e$ have a unique global minimum within the parameter space for finite
samples, and the theory does not offer much guidance about how to deal with a
multiplicity of local minima having similar $e^{T}e$ values that deviate
within sampling error. \ The minimum with the lowest observed $e^{T}e$ value
need not be the closest minimum to the truth.

\begin{figure}
[ptb]
\begin{center}
\includegraphics[
natheight=3.986in,
natwidth=8.5in,
height=132.75pt,
width=280.375pt
]%
{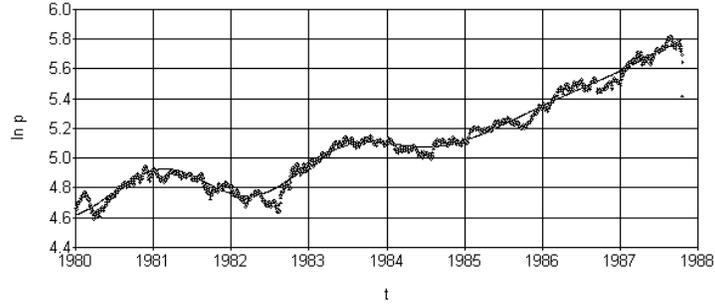}%
\caption{Fit of Eq. (\ref{logps}) to the S\&P 500 from 1980 to 1987 with
$\beta=1.06$. \ Parameters are given in the second column of Table
\ref{parval}.}%
\label{sp8087b1}%
\end{center}
\end{figure}

\qquad The first and second columns of Table \ref{parval} describe two such
local minima for the years prior to the 87 crash. \ The corresponding fits are
plotted in Figs.~\ref{8087b.7} and \ref{sp8087b1}. \ Notice that in the fit of
Fig. \ref{8087b.7}, $\beta=.74$ in accord with the microscopic model of
\cite{JLS 98}. \ However, the fit of Fig. \ref{sp8087b1} has $\beta=1.06$, and
the square of the residual vector is a bit smaller in this second fit. \ 

\qquad In addition to the multiplicity issue, we face another difficulty
because there is really no question that the $p_{t}$ are serially correlated,
which complicates the estimation of standard errors even for the linear
parameters $A$, $B$, and $C$. \ As a result, we do not report standard errors
in Table \ref{parval}, and we will postpone a discussion of whether the
individual terms in Eq. (\ref{logps}) are statistically significant until the
next section, where we consider the first differences of the price index.%

\begin{table}[tbp] \centering
\begin{tabular}
[c]{ccccccc}%
$t^{\ast}$ & 10/19/87 & 10/19/87 & 10/3/74 & 12/2/85 & 10/27/97 & 8/31/98\\
$n$ & 1959 & 1959 & 342 & 365 & 397 & 352\\
$t_{1}$ & 1/2/80 & 1/2/80 & 4/26/73 & 5/23/84 & 3/7/96 & 3/11/97\\
$t_{n}$ & 9/30/87 & 9/30/87 & 8/30/74 & 10/31/85 & 9/30/97 & 7/31/98\\
$t_{c}$ & 11/14/87 & 9/30/87 & 9/22/74 & 11/14/85 & 1/3/98 & 3/2/99\\
$A$ & 5.89 & 5.77 & 2.84 & 5.24 & 7.07 & 7.29\\
$B$ & -.0046 & -.00041 & 1.13 & $-2\times10^{-10}$ & -.0066 & -.0024\\
$C$ & .00051 & -.000056 & .0199 & $4\times10^{-11}$ & -.00042 & .00022\\
$\beta$ & 0.74 & 1.06 & 0.10 & 3.53 & .73 & .84\\
$\omega$ & 9.64 & -2.24 & -1.20 & 14.64 & 4.95 & 10.41\\
$\phi$ & -4.34 & -0.38 & -0.02 & -6.47 & 1.60 & -10.58\\
$\Delta_{t}$ & 8.39 yr & 4.84 yr & 489.63 yr & 0.98 yr & 3.18 yr & 12.74 yr\\
$\Delta_{\omega}$ & 11.65 & 28.85 & 55.90 & -0.30 & 31.63 & 72.44\\
$e^{T}e$ & 4.02 & 3.36 & .24 & .12 & .18 & 0.15\\
$R^{2}$ & .978 & .981 & .932 & .944 & .973 & .969
\end{tabular}
\caption{Parameter values of fits to $\ln(p)$ for the logarithmic
specification\label{parval}}%
\end{table}%

\begin{figure}
[ptb]
\begin{center}
\includegraphics[
natheight=3.986in,
natwidth=8.5in,
height=132.75pt,
width=280.375pt
]%
{lpfig2.epsi}%
\caption{Fit of Eq. (\ref{logps}) to the S\&P 500 from 1980 to 1987 with
$\beta=.74$. \ Parameters are given in the first column of Table
\ref{parval}.}%
\label{8087b.7}%
\end{center}
\end{figure}

\begin{figure}
[ptb]
\begin{center}
\includegraphics[
natheight=4.694in,
natwidth=8.625in,
height=150.8125pt,
width=275pt
]%
{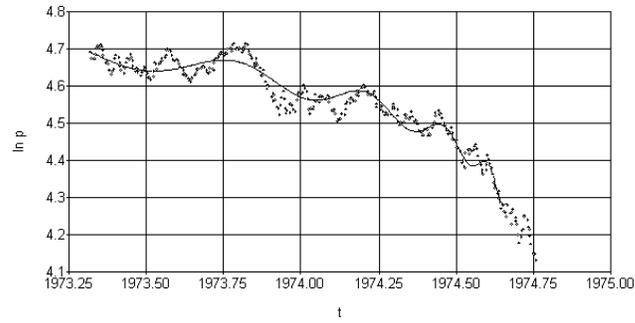}%
\caption{Fit of Eq. (\ref{logps}) to the S\&P 500 from April 1973 to October
1974. \ Parameters are given in the third column of Table \ref{parval}.}%
\label{sp7374}%
\end{center}
\end{figure}

\begin{figure}
[ptb]
\begin{center}
\includegraphics[
natheight=4.188in,
natwidth=7.725in,
height=150.8125pt,
width=276.125pt
]%
{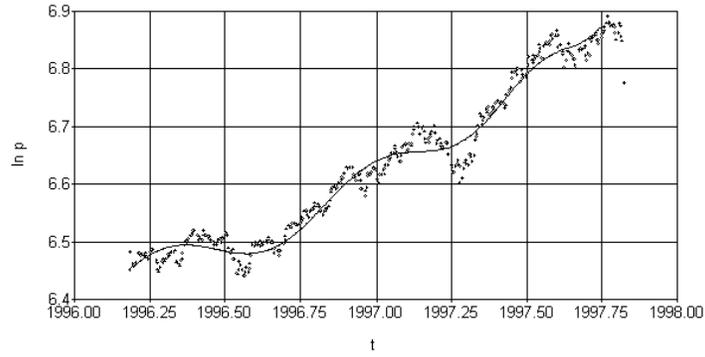}%
\caption{Fit of Eq. (\ref{logps}) to the S\&P 500 from March 1996 to October
1997. \ Parameters are given in the fifth column of Table \ref{parval}.}%
\label{sp9697}%
\end{center}
\end{figure}

\begin{figure}
[ptb]
\begin{center}
\includegraphics[
natheight=4.3in,
natwidth=7.788in,
height=150.8125pt,
width=271pt
]%
{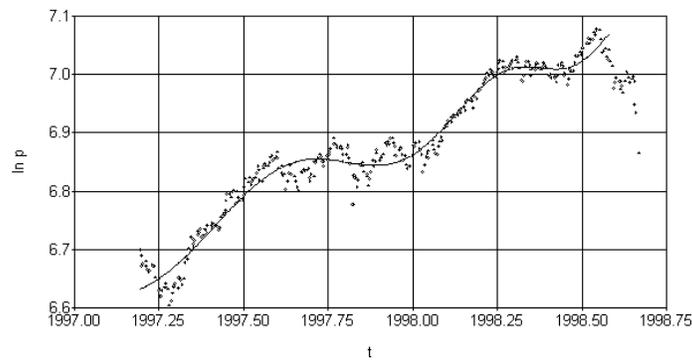}%
\caption{Fit of Eq. (\ref{logps}) to the S\&P 500 from March 1997 to August
1998. \ Parameters are given in the last column of Table \ref{parval}.}%
\label{sp9798}%
\end{center}
\end{figure}

\begin{figure}
[ptb]
\begin{center}
\includegraphics[
natheight=4.275in,
natwidth=7.825in,
height=150.75pt,
width=273pt
]%
{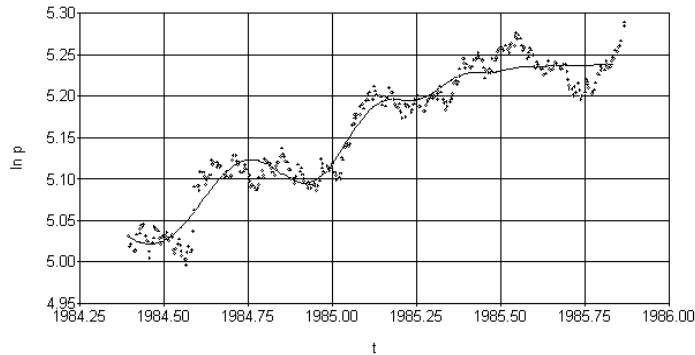}%
\caption{Fit of Eq. (\ref{logps}) to the S\&P 500 from May 1984 to December
1985. \ Parameters are given in the fourth column of Table \ref{parval}.}%
\label{sp8485}%
\end{center}
\end{figure}

\qquad For now, we will simply observe that it is possible to obtain a
``reasonable'' fit of Eq. (\ref{logps}) to the S\&P 500 between 1980 and 1987.
\ Between 1962 and 1998, the next three most significant drawdowns occurred in
October 1974, October 1997, and August 1998. \ As we demonstrate in Figs.
\ref{sp7374}-\ref{sp9798}, ``reasonable''\ fits (as reasonable as the fits in
\cite{JLS 99} for example) to Eq. (\ref{logps}) can also be obtained during
the year prior to each of these drops. \ This evidence might lead one to
conclude that there must be a correlation between spells of log-periodic
fluctuation and financial crashes.

\qquad However, this would be fallacious reasoning because we have not
demonstrated that such spells do not occur (with significant frequency) during
periods that do not culminate in a financial crash. \ In fact, when we
randomly selected a drawdown from the list of drawdowns between 1962 and 1998
with total magnitude on the order of 1\%, we picked a drop of 0.84\% that
occurred at the beginning of December 1985. \ We display the S\&P 500 during
the year prior to this small drop in Fig. \ref{sp8485}, and we see that this
also can be fit to Eq. (\ref{logps}).

\qquad This suggests the possibility that there is nothing particularly
special about a finding of log-periodic fluctuations in the stock market.
\ Laloux et al. \cite{Laloux 99} also relate an experience where they observed
log-periodic fluctuations in a financial index, predicted a crash, and then
were disappointed when none occurred. \ The hypothesis of a correlation may
still be rescued, however, if it can be shown that fits with parameter values
in a select set are harbingers of crashes.

\qquad This was the approach taken by Feigenbaum and Freund \cite{FF 96},
although their definition of a log-periodic spell was ad hoc and not motivated
by any statistical theory. \ Between 1980 and 1994, they found three
independent periods that satisfied their definition. \ A Monte Carlo
simulation extending over 10,000 days, which modeled the stock market as a
random walk, produced 17 such spells. \ The average time between a
log-periodic spell and the next crash was substantially less in the empirical
data than it was in the simulation, which suggests that there might indeed be
an association between log-periodic crashes and financial crashes.\ \ It
should be noted that they worked with the specification of Eq. (\ref{lp}),
which has fewer parameters than the specification of Eq. (\ref{logps}). \ It
is possible that adding these extra parameters makes ``good'' fits to Eq.
(\ref{logps}) less extraordinary.

\qquad Johansen et al. (\cite{JLS 98}, \cite{JLS 99}) describe a similar
experiment where they modeled the stock market as a GARCH process and searched
for fits which extended for about eight years and had $\alpha$, $\omega$, and
$\Delta_{t}$ parameters that fell within a specific interval. \ They also did
not find any linkage between a log-periodic spell and an ensuing crash.
\ Another test on the statistical significance of log-periodicity was done by
Huang et al. \cite{HJLSS 00} in the context of earthquakes. \ They studied the
Lomb periodograms of Monte Carlo simulations generated by a DGP with normal
i.i.d. noise in order to assess the statistical significance of observed periodograms.

\qquad Unfortunately, all of these results are weakened by the fact that only
one possible data generating process was ruled out by each experiment.
\ Furthermore, the DGPs used in the two finance experiments were chosen for
their simplicity, not because they are credible models of the actual DGP for a
stock price . \ Feigenbaum and Freund used a random walk, but there is really
no question any more in the finance community that stock prices do not follow
random walks. \ As for GARCH processes, while these are considered a much
better model of the stock market during ordinary periods, there is no reason
to believe they are an appropriate model of the stochastic process during an
extreme event like a financial crash. \ In fact, the frequency of crashes in
Johansen et al.'s\ Monte Carlo simulations was much smaller than the frequency
of crashes in real data, so their DGP obviously does not adequately capture
the behavior of stock prices during and leading up to a crash.

\qquad Both Feigenbaum and Freund, and Sornette et al. also looked at randomly
selected time windows in the real data and generally found no evidence of
log-periodicity in these windows unless they were looking at a time period in
which a crash was imminent. \ While these results are not tied to any DGP
since there is no Monte Carlo simulation involved, it is still not clear what
if any conclusions can be drawn from this result. \ It is impossible to
collect a random sample of any size with independent draws from a single
century of data. \ Perhaps in the year 3000, our descendants will be better
able to judge by this method if log-periodicity really is an indicator of
financial crashes. \ For those of us who do not expect to live that long, some
more clever approach must be invented if we are to answer this question with
any degree of confidence.\bigskip

\section{Behavior of First Differences\label{FD}\bigskip}

\qquad If we focus on a specific choice of the nonlinear parameters $\beta$,
$\omega$, $\phi$, $t_{c}$, $\Delta_{t}$, and $\Delta_{\omega}$ in the
specification of Eq. (\ref{logps}), then we have a specification linear in the
parameters $A$, $B$, and $C$, and we could naively perform statistical tests
on hypotheses regarding the values of these parameters. \ For example, it
would be of great interest to determine if $C$, the coefficient of the term
responsible for producing log-periodic oscillations in Eq. (\ref{logps}) is
significantly different from zero.

\qquad Unfortunately, a minor obstacle gets in the way of our doing this in a
consistent manner. \ Our hypothesis is that if $j(t)=0$, then
\begin{equation}
y(t)=\ln(p(t))=A+Bf_{1}(t)+Cf_{2}(t)+\varepsilon(t),\label{simpspec}%
\end{equation}
where $\varepsilon(t)$ is a random disturbance term satisfying
\begin{equation}
E_{t_{0}}[\varepsilon(t)]=0.
\end{equation}
There are well-known ordinary least squares (OLS) estimation formulas both for
point estimates of $A$, $B$, and $C,$ and their corresponding standard errors.
\ These formulas apply if the $\varepsilon(t)$ are homoskedastic and
disturbances at different times are uncorrelated. \ However, if we consider
the S\&P\ 500 from January 1980 to September 1987, the lag-1
autocorrelation\footnote{We refer to the correlation between $\varepsilon
(t_{i})$ and $\varepsilon(t_{i-l})$ as the lag-$l$ autocorrelation.} for
$\ln(p(t))$ is .9994, which is not surprising since security prices were once
believed to follow random walks. \ Stock prices may move up or down on a given
day, but they rarely move far from the previous day's closing.

\qquad We can eliminate most, if not all, of this serial correlation by
looking not at $\ln(p(t))$ but at the first differences of $\ln(p(t))$.
\ Define
\begin{equation}
y_{i}=\ln(p(t_{i}))
\end{equation}
and
\begin{equation}
\Delta y_{i}=\ln(p(t_{i}))-\ln(p(t_{i-1}))\text{.}%
\end{equation}
Then the lag-1 autocorrelation for $\Delta y_{i}$ is only .1511 for the same
period examined above.

\qquad Let us also define
\begin{align}
\Delta t_{i}  & =t_{i}-t_{i-1}\\
\Delta f_{1i}  & =f_{1}(t_{i})-f_{1}(t_{i-1})\\
\Delta f_{2i}  & =f_{2}(t_{i})-f_{2}(t_{i-1})\text{.}%
\end{align}
Then Eq. (\ref{simpspec}) implies
\begin{equation}
E_{t_{0}}[\Delta y_{i}|j(t_{i})=0]=B\Delta f_{1i}+C\Delta f_{2i}%
\text{.}\label{increment}%
\end{equation}
We will consider the augmented specification
\begin{equation}
E_{t_{0}}[\Delta y_{i}|j(t_{i})]=A+B\Delta f_{1i}+C\Delta f_{2i}+D_{2}%
\delta_{2i}+D_{3}\delta_{3i}+D_{4}\delta_{4i}.\label{regress}%
\end{equation}
The additional variables $\delta_{2i}$, $\delta_{3i}$, and $\delta_{4i}$ are
binary indicator variables\ (or so-called ``dummy variables'') which take on
values of 0 or 1. \ For a given $s$, the variable $\delta_{si}$ will equal 1
if $\Delta t_{i}$ equals $s$ and 0 otherwise. \ The inclusion of the constant
term allows for the possibility of a constant drift in the stock price index.
\ Since the magnitude of this drift may depend on the length of time between
measurements, we also include the dummy variables. \ (Four is the largest
$\Delta t$ that appears in the data.) \ By doing this, we can, for example,
correct for weekend effects. \ These are of concern because weekly cycles in
$\Delta y$ might arise naturally from the institutional details of the market,
and, if we did not include them in the regression, an oscillatory function
like $\Delta f_{2}$ might easily project onto them, biasing the estimate of
its coefficient.%

\begin{figure}
[ptb]
\begin{center}
\includegraphics[
natheight=4.713in,
natwidth=7.1in,
height=150.75pt,
width=225.125pt
]%
{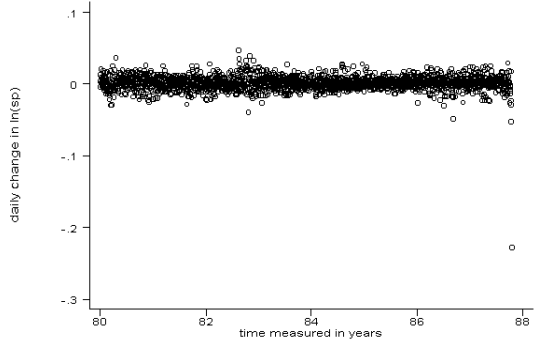}%
\caption{First differences of the logarithm of the S\&P 500 from January 1980
to September 1987.}%
\label{lspdiff}%
\end{center}
\end{figure}

\qquad In Fig. \ref{lspdiff}, we plot $\Delta y_{i}$ as a function of time.
\ No obvious oscillatory pattern is apparent in the graph, although this does
not prove that no pattern is there.%

\begin{table}[tbp] \centering
\begin{tabular}
[c]{lll}%
& 1/2/80-9/30/87 & 1/2/80-6/5/86\\
$A$ & .000668 & .000450\\
$B$ & -.006384 & -.019967\\
$C$ & -.001149 & .001542\\
$D_{2}$ & .001712 & .001646\\
$D_{3}$ & -.002029 & -.002149\\
$D_{4}$ & -.002826 & -.003463\\
$t_{c}$ & 10/7/87 & 6/19/86\\
$\beta$ & 0.590 & 0.497\\
$\omega$ & 7.333 & 8.255\\
$\phi$ & 0.338 & 0.004\\
$\Delta_{t}$ & 13.4 yr & 1285 yr\\
$\Delta_{\omega}$ & 28.80 & -0.79
\end{tabular}
\caption{Parameter values for fits of the first differences of the S\&P 500
from January 1980 to September 1987 using the log-periodic specification.\label
{parval2}}%
\end{table}%

\begin{table}[tbp] \centering
\begin{tabular}
[c]{lllll}%
$n$ & 1958 &  &  & \\
$e^{T}e$ & .159777 &  &  & \\
$R^{2}$ & 0.0172 &  &  & \\
&  &  &  & \\
& \multicolumn{1}{||l}{coeff.} & std. err. & $T$ & $p$\\\hline\hline
$A$ & \multicolumn{1}{||l}{.000665} & .000273 & 2.434 & .015\\
$B$ & \multicolumn{1}{||l}{-.006334} & .004826 & -1.312 & .190\\
$C$ & \multicolumn{1}{||l}{-.001130} & .000251 & -4.504 & .000\\
$D_{2}$ & \multicolumn{1}{||l}{.001704} & .002153 & 0.791 & .429\\
$D_{3}$ & \multicolumn{1}{||l}{-.002034} & .000605 & -3.359 & .001\\
$D_{4}$ & \multicolumn{1}{||l}{-.002832} & .001450 & -1.953 & .051
\end{tabular}
\caption{Estimation of linear parameters to log-periodic specification with
full data set.\label{estlinu}}%
\end{table}%

\qquad Using the same downward simplex method used in Section \ref{logspec},
we obtain the parameter values listed in the first column of Table
\ref{parval2}. \ In Table \ref{estlinu}, we report the results of the
corresponding OLS regression, holding the nonlinear parameters fixed (the
linear parameters deviate slightly from Table \ref{parval2} because they are
computed by a different program). \ In addition to OLS standard errors, we
report corresponding $T$ statistics and nominal $p$ values which correspond to
the probability that $\left|  T\right|  $ will be greater than or equal to its
observed value under a null hypothesis that the given coefficient is zero.
\ Note that these nominal $p$ values are computed under the assumption of
homoskedasticity and no serial correlation. \ It is fairly well accepted that
the S\&P 500 is a heteroskedastic process. \ Furthermore, while the
autocorrelation of $\Delta y_{t}$ may be insignificant, it is well known that
$\left|  \Delta y_{t}\right|  $ typically shows significant autocorrelation,
so the $\Delta y_{t}$ time series is very likely serially dependent.
\ Therefore the reported $p$ values should not be given much credence.

\qquad The search for standard error estimators which consistently estimate
the standard deviation of regression coefficients, even for a model which
exhibits heteroskedasticity and autocorrelation of an unknown nature, is
presently an active field of econometric research. \ We explored both the
Newey-West (see Chapters 11-12 of \cite{Greene 00}) and the Kiefer-Vogelsang
\cite{KV 00} approaches. However, our results were not consistent between
these different methods, which suggests that our sample size is too small for
the asymptotic probability theory underlying these methods to be a good
approximation. \ Consequently, it proved necessary to use Monte Carlo
simulations to estimate the finite-sample probability distribution of the $T$ statistics.

\qquad In order to perform Monte Carlo simulations, the null hypothesis must
completely specify the data-generating process (DGP) to be simulated. \ The
null hypothesis in this context, which is just the blanket assumption that
there is no log-periodicity, is a compound hypothesis and does not fully
specify a DGP. \ However, our main concern is whether any apparent
log-periodicty can be accounted for by a simple specification of the index's
behavior such as a random walk or a GARCH\ process. \ Matching the first and
second moments of $\Delta y$ conditional upon $\Delta t$, we constructed a
random walk with normally distributed innovations.

\qquad In 20 simulations, six produced fits to the specification of
(\ref{regress}) with $T$ statistics for $B$ larger in magnitude than the
observed value of -1.312, which implies a $p$ value of .30$\pm.10$. \ Thus the
$\Delta f_{1}$ term is not statistically significant. \ This is not entirely
surprising since the inclusion of the $f_{1}$ term in Eq. (\ref{simpspec}) is
really not well motivated in \cite{SJ 97}. \ Unlike the $f_{2}$ term, the
$f_{1}$ term does not come out of Sornette and Johansen's Landau expansion.
\ Rather, it is included ad hoc to account for the growth trend seen in most
financial indices, though this growth trend could also be accounted for by a
constant drift. \ On the other hand, we observe only one simulation with a $T
$ statistic for the log-periodic coefficient $C$ as large as 4.5 in magnitude.
\ This implies a $p$-value of .050$\pm.049$, so we do not firmly reject the
log-periodic term at the 5\% level.

\qquad If the Sornette-Johansen paradigm has any practical value though, we
should be able to predict a crash even if we do not have all the data leading
right up to the crash. \ What happens if we disregard the last year of stock
market data?\ \ Do the first differences of the S\&P\ 500 contain the
information necessary to forecast a crash in October of 1987?%

\begin{table}[tbp] \centering
\begin{tabular}
[c]{lllll}%
$n$ & 1624 &  &  & \\
$e^{T}e$ & .128761 &  &  & \\
$R^{2}$ & 0.0143 &  &  & \\
&  &  &  & \\
& \multicolumn{1}{||l}{coeff.} & std. err. & $T$ & $p$\\\hline\hline
$A$ & \multicolumn{1}{||l}{.000451} & .000309 & 1.458 & .145\\
$B$ & \multicolumn{1}{||l}{-.020201} & .009796 & -2.062 & .039\\
$C$ & \multicolumn{1}{||l}{.001555} & .000481 & 3.232 & .001\\
$D_{2}$ & \multicolumn{1}{||l}{.001647} & .002327 & 0.708 & .479\\
$D_{3}$ & \multicolumn{1}{||l}{-.002148} & .000678 & -3.169 & .002\\
$D_{4}$ & \multicolumn{1}{||l}{-.003461} & .001593 & -2.172 & .030
\end{tabular}
\caption
{Estimation of linear parameters to the best log-periodic specification for
the truncated data set.\label{regress86}}%
\end{table}%

\qquad Using a window of data from January 1980 to June 1986, the best fit we
obtain is reported in the second column of Table \ref{parval2}, and the
corresponding OLS regression is given in Table \ref{regress86}. \ Notice that
we do not forecast a crash in 1987. \ Rather, we forecast a crash just beyond
the end of the window. \ Using the same DGP as before, in five of 20
simulations we obtained a $T$ statistic larger than 3.23 for the log-periodic
term, implying $p=.250\pm.097$. \ Thus the log-periodic term is not
significant at the 5\% level in this fit.

\qquad As we noted in Section \ref{logspec}, the same data set may give
several fits with roughly similar values of $e^{T}e$. \ Although the nonlinear
parameters of the fit for the full data set do not give the best fit for the
truncated data set, conceivably they could still produce a decent fit with a
statistically significant log-periodic component. \ An OLS regression using
the nonlinear parameters of the first column of Table \ref{parval2} is
reported in Table \ref{regress87b}. \ In 20 simulations, 14 had $T$ statistics
for $C$ larger in magnitude than 2.767, so $p=.700\pm.102$. \ Thus even using
the nonlinear parameters that were statistically significant for the full data
set, we do not find a statistically significant log-periodic component in the
truncated data set.%

\begin{table}[tbp] \centering
\begin{tabular}
[c]{lllll}%
$n$ & 1624 &  &  & \\
$e^{T}e$ & .129005 &  &  & \\
$R^{2}$ & 0.0124 &  &  & \\
&  &  &  & \\
& \multicolumn{1}{||l}{coeff.} & std. err. & $T$ & $p$\\\hline\hline
$A$ & \multicolumn{1}{||l}{.000065} & .000438 & 0.147 & .883\\
$B$ & \multicolumn{1}{||l}{-.032800} & .016639 & -1.971 & .049\\
$C$ & \multicolumn{1}{||l}{-.000879} & .000318 & -2.767 & .006\\
$D_{2}$ & \multicolumn{1}{||l}{.000723} & .002362 & 0.306 & .759\\
$D_{3}$ & \multicolumn{1}{||l}{-.002992} & .000924 & -3.239 & .001\\
$D_{4}$ & \multicolumn{1}{||l}{-.004185} & .001829 & -2.288 & .022
\end{tabular}
\caption
{Estimation of linear parameters for a log-periodic specification to the truncated
data set using the best nonlinear parameters obtained for the full data set.
\label{regress87b}}%
\end{table}%

\qquad Since the log-periodic function is significant for the full data set,
there must be some structure in the data which projects onto this function.
\ However, the failure of this function to produce a significant coefficient
in the truncated data set suggests the pertinent structure is localized within
the last year before the crash. \ Note that this structure need not be
log-periodic. \ We can only conclude that the log-periodic function represents
this structure better than any of the other functions included in the regression.

\qquad Whatever is happening in the last year, we can conclude that $E[\Delta
y]$ does not have a statistically significant log-periodic trend through most
of the data set. \ More precisely, the $\Delta f_{2}$ term in the
specification of Eqs. (\ref{increment}) and (\ref{regress}) is not
significant. \ Therefore, the $f_{2}$ term in Eq. (\ref{simpspec}) is not
significant, which would imply that $E[\ln p_{t}]$ has no log-periodic trend.

\qquad One might wonder how this can be reconciled with the fits observed in
Section \ref{logspec}. \ How can $\ln(p)$ behave log-periodically when its
expectation does not? \ Serial correlation may answer this question. \ A short
string of several increments which coincidentally fall in the same direction
followed by a string of increments in the opposite direction may create an
oscillation completely by happenstance with no underlying force behind it.\bigskip

\section{The Frequency Distribution of Drawdowns\label{DDD}\bigskip}

\begin{figure}
[ptb]
\begin{center}
\includegraphics
[
natheight=284.500000pt,
natwidth=426.812500pt,
height=150.75pt,
width=225.125pt]
{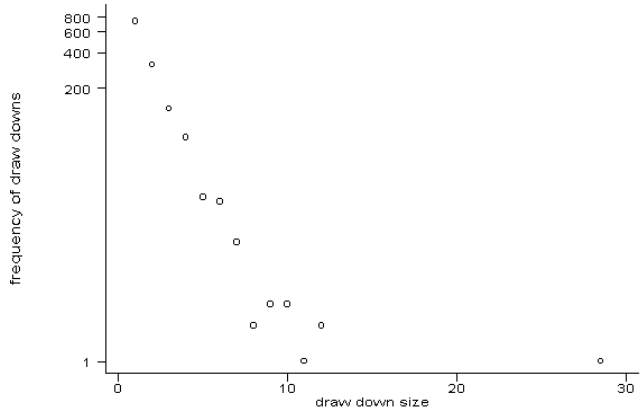}%
\caption{Frequency of draw downs for the S\&P 500 index from 1962 to 1998.
\ The bin size is 1\%.}%
\label{ddf}%
\end{center}
\end{figure}

\qquad In their search for an independent form of evidence to corroborate
their hypothesis that sizable financial crashes are often the culmination of
forces that build up over long periods of time, Johansen and Sornette \cite{JS
99b} examined the frequency distribution of drawdowns, a drawdown being the
cumulative drop in a financial index from a local maximum to the succeeding
local minimum. \ Fig. \ref{ddf} displays a semilog graph of the frequency
distribution of drawdowns larger than 0.5\% for the S\&P 500 during the period
1962-1998. \ If we disregard the lone drawdown of magnitude larger than 20\%,
which corresponds to the October 1987 crash, it is immediately apparent that
this semilog graph looks linear. \ Seizing upon this observation, Sornette and
Johansen concluded that there must be two populations of drawdowns. \ The
majority of drawdowns have an exponential distribution, and these constitute
the first population. \ Those drawdowns, like the 87 crash, which do not obey
this exponential distribution constitute the second population. \ Sornette and
Johansen further suggest that the first population is produced by the normal
behavior of a financial index obeying a GARCH process. \ In contrast, the
second population is the result of the imitative behavior of investors, and it
is this population that exhibits log-periodic oscillations prior to the drawdown.%

\qquad While the frequency distribution of the first population of drawdowns
(every data point except the 87 outlier) may look exponential, we can
objectively test whether it actually is exponential. \ In Table \ref{ddr}, we
present the results of a regression of the logarithm of the frequency versus a
quadratic function of the drawdown size $d$. \ The alleged outlier, the
October 87 crash, was not included as a data point in this regression.
\ Nevertheless, the quadratic term in this regression with the 87 crash
excluded is statistically significant at the 1\% level. \ The justification
for treating the 87 crash separately was that it did not fit to a line through
the other drawdowns. \ However, the significance of the quadratic term
indicates the logarithm of the frequency distribution for the remaining
drawdowns is not linear either. \ Since a nonexponential frequency
distribution can certainly be found that fits to all the data, the visual
evidence of two populations is unconvincing.%

\begin{table}[tbp] \centering
\begin{tabular}
[c]{ccccc}%
\# of obs. & 12 &  &  & \\
$e^{T}e$ & .180 &  &  & \\
$R^{2}$ & .969 &  &  & \\
&  &  &  & \\
& \multicolumn{1}{||c}{coefficient} & standard error & $T$ & $p$\\\hline\hline
$d$ & \multicolumn{1}{||c}{-1.097} & 0.155 & -7.065 & 0.000\\
$d^{2}$ & \multicolumn{1}{||c}{0.039} & 0.012 & 3.378 & 0.008\\
constant & \multicolumn{1}{||c}{7.812} & 0.439 & 17.801 & 0.000
\end{tabular}
\caption{Regression of the logarithm of the frequency of draw downs versus a
quadratic function of the draw down size $d$.  The October 1987 crash is not included.
\label{ddr}}%
\end{table}%

\qquad Aware of these deviations from the exponential curve, Johansen and
Sornette have recently proposed (\cite{JS 00}) that the first population is
better described by a ``stretched'' exponential
\begin{equation}
N(d)=\exp(A+Bd^{x}),\label{stretched}%
\end{equation}
where $N(d)$ is the frequency of drawdowns of size $d$. \ They find that
$x\approx.8-.9$. \ Using this specification on our data set, we find the best
value of $x$ within the interval $[.8,1]$ to be .8 and obtain the regression
estimates of Table \ref{ddr2}. \ Notice that the $d^{2}$ term remains
statistically significant at the 5\% level. \ Thus the specification of the
stretched exponential with $x\in\lbrack.8,1]$ must also be rejected.%

\begin{table}[tbp] \centering
\begin{tabular}
[c]{ccccc}%
\# of obs. & 12 &  &  & \\
$e^{T}e$ & .196 &  &  & \\
$R^{2}$ & .9666 &  &  & \\
&  &  &  & \\
& \multicolumn{1}{||c}{coefficient} & standard error & $T$ & $p$\\\hline\hline
$d^{.8}$ & \multicolumn{1}{||c}{-1.549} & 0.230 & -6.722 & 0.000\\
$d^{2}$ & \multicolumn{1}{||c}{0.022} & 0.010 & 2.280 & 0.049\\
constant & \multicolumn{1}{||c}{8.348} & 0.534 & 15.640 & 0.000
\end{tabular}
\caption{Regression of the logarithm of the frequency of draw downs versus
$d^{.8}$ and $d^2$, where $d$ is the draw down size $d$.
The October 1987 crash is not included. \label{ddr2}}%
\end{table}%

\qquad However, if we allow $x$ to range beyond this interval, we find that
$x=.5$ gives the best fit, and if we include a term linear or quadratic in $d$
neither is statistically significant at the 5\% level. \ Furthermore, if we
consider a specification of the form
\begin{equation}
N(d)=\exp(A+B\sqrt{d}+Cd),
\end{equation}
then we are able to accomodate all the data including the 87 outlier. \ In
Table \ref{ddr3}, we give the regression estimates for such a specification
and we obtain a favorable $R^{2}=.9519$, so with this data set it does not
seem reasonable at all to treat the 87 crash differently from other drawdowns.
\ Nevertheless, it is possible that this result is an artifact of the small
data set available to us.

\qquad Johansen and Sornette considered the specification (\ref{stretched})
over the distribution of drawdowns for a much larger time period. \ They also
looked at many other financial time series besides the S\&P 500, and they
consistently obtained values of $x\sim1$. \ In more than half of these
examples, though, it is patently obvious that the specification
(\ref{stretched}) is wrong, for statistically significant deviations are
clearly visible in their graphs. \ If the model is misspecified, no valid
inferences can be drawn from it.%

\begin{table}[tbp] \centering
\begin{tabular}
[c]{ccccc}%
\# of obs. & 13 &  &  & \\
$e^{T}e$ & .289 &  &  & \\
$R^{2}$ & .9519 &  &  & \\
&  &  &  & \\
& \multicolumn{1}{||c}{coefficient} & standard error & $T$ & $p$\\\hline\hline
$d^{.5}$ & \multicolumn{1}{||c}{-5.148} & 0.570 & -9.023 & 0.000\\
$d$ & \multicolumn{1}{||c}{0.542} & 0.089 & 6.069 & 0.000\\
constant & \multicolumn{1}{||c}{11.847} & 0.842 & 14.071 & 0.000
\end{tabular}
\caption{Regression of the logarithm of the frequency of draw downs versus
$d^{.5}$ and $d$, where $d$ is the draw down size $d$.
The October 1987 crash is included. \label{ddr3}}%
\end{table}%

\qquad Of course, in Section \ref{logspec}, we saw that the 87 crash was not
the only drawdown between 1962 and 1998 with a log-periodic precursor. \ If we
had an objective way of distinguishing drawdowns with log-periodic precursors
from drawdowns without such precursors, then it is still possible that the
distribution of drawdowns without precursors is exponential.\bigskip

\section{Concluding Remarks\label{CR}\bigskip}

\qquad While the evidence of log-periodic precursors to financial crashes has
spawn\-ed a minor industry in the physics community, it is apparent that a
more rigorous investigation of this phenomenon must be conducted before we can
come to a firm conclusion that log-periodic oscillations reflect a discrete
scale invariance underlying the economy and furthermore that such oscillations
signal an impending crash. \ In particular, it would be desirable for
researchers to agree upon an objective definition of what constitutes a
log-periodic spell so that a statistical association between these spells and
ensuing crashes can be firmly established.

\qquad We must emphasize that a physicist who endeavors to successfully carry
out these goals will have to employ and perhaps even invent advanced
statistical methods of a nature that may be unfamiliar to him because they are
simply unnecessary in more customary physics settings. \ A physicist can
typically rely more confidently on elementary statistical methods because he
can do controlled experiments. \ He can often identify and directly measure
the relevant sources of error. \ Testing a model can be just a matter of
determining whether the model agrees with the data to within a degree that can
be accounted for by these known sources of error. \ In contrast, an
econometrician's task is vastly more difficult, at least as far as testing
models goes, for he must simultaneously estimate the parameters of his model
and the properties of any unmodeled disturbances using the same data set
(which was probably gathered under less than ideal conditions). \ As a result,
econometricians have had to acquire complicated tools to efficiently salvage
as much information as they can from their data, and physicists who expand
into this realm will have to do the same.

\begin{figure}
[ptb]
\begin{center}
\includegraphics[
natheight=4.861in,
natwidth=8.542in,
height=138.0625pt,
width=273.75pt
]%
{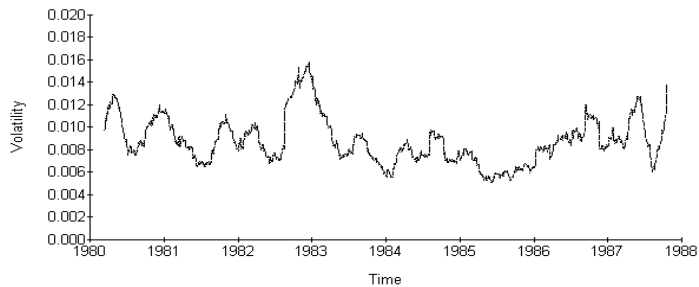}%
\caption{The standard deviation of daily changes to the logarithm of the
S\&P~500 sampled over a moving window of fifty trading days from January 1980
to October 1987.}%
\label{volatility}%
\end{center}
\end{figure}

\qquad While the present work does not refute what has come before, we think
it does shed some doubt on the hypothesis of discrete scale invariance in
financial markets. \ Though log-periodicity is visually apparent in the S\&P
500 preceeding the 87 crash, we find no evidence of log-periodicity in the
mean function of this index prior to a year before the crash. \ This would
suggest that if log-periodicity is embedded in the stochastic process of the
index, then it must be conveyed in higher-order moments. \ However, evidence
of log-periodicity in the second moment of the process is also lacking. \ In
Fig. \ref{volatility}, we plot a measure of the volatility of the S\&P 500,
the standard deviation of $\Delta y$ sampled over a moving window of fifty
trading days, from January 1980 to October 1987. \ No pattern of
log-periodicity can be seen in the graph.%

\qquad The negative result of this paper is also troubling from a theoretical
perspective. \ While Johansen et al. (\cite{JLS 98}) are not alone in their
efforts to construct an economic model which can produce log-periodic price
series (see for example \cite{Canessa 00}), economists would be reluctant to
embrace any theory that diverges from Johansen et al.'s attempt to ensure the
model is robust to the introduction of a rational agent with rational
expectations. \ There are today economic theorists exploring models in which
agents are not fully rational, but a model whose predictions absolutely depend
on there being no one in the economy who has a good grasp of how the world
works is not likely to acquire a large following. \ Any model which does not
suffer from this disease will have to satisfy a restriction like the
martingale condition of Eq.~(\ref{gen mar2}).

\qquad A model which incorporates discrete scale invariance and
log-periodicity in its structure would not necessarily imply that the
expectation of prices must also behave log-periodically. \ However, if
log-periodicity arises because the probability of a crash behaves
log-periodically, then it is difficult to imagine how one could avoid making
the prediction that the expected price must also behave log-periodically. \ If
nothing else, the results of this paper would seem to demonstrate that a more
circuitous explanation of log-periodicity is required.\bigskip

\noindent{\Large \textbf{Acknowledgements\bigskip} }

{\normalsize \qquad We would like to thank William Brock, Peter Freund, Joel
Horowitz, Kirill Ilinski, Sok Bae Lee, Gene Savin, and Robert Wyckham for
comments, discussions, and advice.\bigskip}

\end{document}